\documentclass[sigconf,natbib=false]{acmart}
\usepackage{filecontents}

\usepackage{comgipp}
\usepackage{booktabs} % For formal tables

\usepackage[
	backend=biber,
	style=numeric,
	firstinits=true,
	url=false,
	isbn=false,
	]{biblatex}
\usepackage{qtree}
\usepackage{ifthen}
\usepackage{lstomdoc}
\usepackage{enumitem}
\setlist{leftmargin=5mm}
\usepackage{xfrac}

\usepackage{balance} % for balancing left-right columns
\AtEveryBibitem{%
\clearname{translator}%
\clearfield{pagetotal}%
\clearfield{publisher}%
\clearfield{issn}%
\clearfield{groups}%
\clearfield{volume}
\clearfield{number}
\clearname{editor}
}
\newcommand{\theReferenceText}{\fullcite{Schubotz2018}}

\DeclareFieldFormat
[article,inbook,incollection,inproceedings,patent,thesis,unpublished]
{title}{#1}
\setcopyright{usgovmixed}

\copyrightyear{2018}
\acmYear{2018}
\setcopyright{usgovmixed}
\acmConference[JCDL '18]{The 18th ACM/IEEE Joint Conference on Digital Libraries}{June 3--7, 2018}{Fort Worth, TX, USA}
\acmBooktitle{JCDL '18: The 18th ACM/IEEE Joint Conference on Digital Libraries, June 3--7, 2018, Fort Worth, TX, USA}
\acmPrice{15.00}
\acmDOI{10.1145/3197026.3197058}
\acmISBN{978-1-4503-5178-2/18/06}

\newcommand{\PPPcite}[1]{\href{#1}{\textcolor{red}{[?]}}}
\newcommand{\qId}[1]{\href{https://mathmlben.wmflabs.org/#1}{#1}}
\newcommand{\qvar}[1]{\ensuremath{\textcolor{red}{#1}}}

\newcommand{\w}[2]{\href{https://www.wikidata.org/wiki/#1}{#2}}

\begin{document}
\title[Improving the Representation and Conversion of Mathematical Formulae]{Improving the Representation and Conversion of Mathematical Formulae by Considering their Textual Context}

\author{Moritz Schubotz\textsuperscript{1}, Andr\'{e} Greiner-Petter\textsuperscript{1}, Philipp Scharpf\textsuperscript{1}, \\ Norman Meuschke\textsuperscript{1}, Howard S.~Cohl\textsuperscript{2}, Bela Gipp\textsuperscript{1}}

\affiliation{
	\institution{
    \textsuperscript{1}Information Science Group, University of Konstanz, Germany  (first.last@uni-konstanz.de)
    }
}

\affiliation{
	\institution{
    \textsuperscript{2}Applied and Computational Mathematics Division, NIST, U.S.A. (hcohl@nist.gov)
    }
}
\renewcommand{\shortauthors}{M. Schubotz, A. Greiner-Petter, P. Scharpf, et al.}

\begin{abstract}
Mathematical formulae represent complex semantic information in a concise form.
Especially in Science, Technology, Engineering, and Mathematics, mathematical formulae are crucial to communicate information, e.g., in scientific papers, and to perform computations using computer algebra systems. Enabling computers to access the information encoded in mathematical formulae requires machine-readable formats that can represent both the presentation and content, i.e., the semantics, of formulae. Exchanging such information between systems additionally requires conversion methods for mathematical representation formats. We analyze how the semantic enrichment of formulae improves the format conversion process and show that considering the textual context of formulae reduces the error rate of such conversions.
Our main contributions are:
(1) providing an openly available benchmark dataset for the mathematical format conversion task consisting of a newly created test collection, an extensive, manually curated gold standard and task-specific evaluation metrics;
(2) performing a quantitative evaluation of state-of-the-art tools for mathematical format conversions;
(3) presenting a new approach that considers the textual context of formulae to reduce the error rate for mathematical format conversions.
Our benchmark dataset facilitates future research on mathematical format conversions as well as research on many problems in mathematical information retrieval. Because we annotated and linked all components of formulae, e.g., identifiers, operators and other entities, to Wikidata entries, the gold standard can, for instance, be used to train methods for formula concept discovery and recognition. Such methods can then be applied to improve mathematical information retrieval systems, e.g., for semantic formula search, recommendation of mathematical content, or detection of mathematical plagiarism.
\end{abstract}

 \begin{CCSXML}
<ccs2012>
<concept>
<concept_id>10002951.10003317.10003371.10003381.10003383</concept_id>
<concept_desc>Information systems~Mathematics retrieval</concept_desc>
<concept_significance>500</concept_significance>
</concept>
<concept>
<concept_id>10010405.10010497.10010510.10010514</concept_id>
<concept_desc>Applied computing~Format and notation</concept_desc>
<concept_significance>300</concept_significance>
</concept>
</ccs2012>
\end{CCSXML}

\ccsdesc[500]{Information systems~Mathematics retrieval}
\ccsdesc[300]{Applied computing~Format and notation}

\keywords{MathML; goldstandard; dataset; computer algebra systems}

\sloppy
\maketitle
\thispagestyle{firststyle}
\fussy
\section{Introduction} \label{sc.intro}
In STEM disciplines, i.e., Science, Technology, Engineering, and Mathematics, mathematical formulae are ubiquitous and crucial to communicate information in documents, such as scientific papers, and to perform computations in computer algebra systems (CAS). Mathematical formulae represent complex semantic information in a concise form that is independent of natural language. These characteristics make mathematical formulae particularly interesting features to be considered by information retrieval systems.

In the digital libraries context, major information retrieval applications for mathematical formulae include search and recommender systems as well as systems that support humans in understanding and applying mathematical formulae, e.g., by visualizing mathematical functions or providing auto completion and error correction functionality in typesetting and CAS.

However, the extensive, context-dependent polysemy and polymorphism of mathematical notation is a major challenge to exposing the knowledge encoded in mathematical formulae to such systems. The amount of mathematical concepts, e.g., mathematical structures, relations and principles, is much larger than the set of mathematical symbols available to represent these concepts. Therefore, the meaning of mathematical symbols varies in different contexts, e.g., in different documents, and potentially even in the same context. Identical mathematical formulae, even in the same document, do not necessarily represent the same mathematical concepts. Identifiers are prime examples of mathematical polysemy. For instance, while the identifier $E$ commonly denotes energy in physics, $E$ commonly refers to expected value in statistics. 

Polymorphism of mathematical symbols is another ubiquitous phenomenon of mathematical notation. For example, whether the operator $\cdot$ denotes a scalar multiplication or a vector multiplication depends on the type of the elements that the operator is applied to. Opposed to programming languages, which handle polymorphism by explicitly providing type information about objects to the compiler, e.g., to check and call methods offered by the specific objects, mathematical symbols mostly denote such type information implicitly so that they need to be reasoned from the context.

Humans account for the inherent polysemy and polymorphism of mathematical notation by defining context-dependent meanings of mathematical symbols in the text that surrounds formulae, e.g., for identifiers, subscripts and superscripts, brackets, and invisible operators. Without such explanations, determining the meaning of symbols is challenging, even for mathematical experts. For example, reliably determining whether $[a,b]$ represents an interval or the commutator $[a,b] = ab - ba$ in ring theory requires information on whether $[]$ represent the Dirac brackets.

Enabling computers to access the full information encoded in mathematical formulae mandates machine-readable representation formats that capture both the presentation, i.e., the notational symbols and their spacial arrangement, and the content, i.e., the semantics, of mathematical formulae. Likewise, exchanging mathematical formulae between applications, e.g., CAS, requires methods to convert and semantically enrich different representation formats. The Mathematical Markup Language (MathML) allows one to encode both presentation and content information in a standardized and extensible way (cf.~Section \ref{sc.gold}).

Despite the availability of MathML, most Digital Mathematical Libraries (DML) currently exclusively use presentation languages, such as TeX and LaTeX to represent mathematical content. On the other hand, CAS, such as MAPLE, Mathematica or SageMath\footnote{The mention of specific products, trademarks, or brand names is for purposes of identification only. Such mention is not to be interpreted in any way as an endorsement or certification of such products or brands by the National Institute of Standards and Technology, nor does it imply that the products so identified are necessarily the best available for the purpose. All trademarks mentioned herein belong to their respective owners.}, typically use representation formats that include more content information about mathematical formulae to enable computations. The conversion between representation formats entails many conceptual and technical challenges, which we describe in more detail in Section~\ref{sc.brw}.
Despite the availability of numerous conversion tools, the inherent challenges of the conversion process result in a high error rate and often lossy conversion of mathematical formulae in different representation formats.

To push forward advances in research on mathematical format conversion, we make the following contributions, which we describe in the subsequent sections: 
\begin{enumerate}
\item We provide an openly available benchmark dataset to evaluate tools for mathematical format conversion (cf.~Section \ref{sc.gold}). The dataset includes:
\begin{itemize}
\item a new test collection covering diverse research areas in multiple STEM disciplines;
\item an extensive, manually curated gold standard that includes annotations for both presentation and content information of mathematical formulae;
\item tools to facilitate the future extension of the gold standard by visually supporting human annotators; and
\item metrics to quantitatively evaluate the quality of mathematical format conversions.
\end{itemize} 
\item We perform an extensive, quantitative evaluation of state-of-the-art tools for mathematical format conversion and provide an automated evaluation framework that easily allows rerunning the evaluation in future research (cf.~Section \ref{sc.evalI}).
\item We propose a novel approach to mathematical format conversion (cf.~Section \ref{sc.appr}). The approach imitates the human sense-making process for mathematical content by analyzing the textual context of formulae for information that helps link symbols in formulae to a knowledge base, in our case Wikidata, to determine the semantics of formulae.
\end{enumerate}

\section{Background \& Related Work} \label{sc.brw}
In the following, we use the Riemann hypothesis (\ref{eq.rh}) as an example to explain typical challenges of converting different representation formats of mathematical formulae:
\begin{align}
\zeta(s) =0 \Rightarrow \Re s = \frac12 \lor \Im s=0.\label{eq.rh}
\end{align}
We will focus on the representation of the formula in LaTeX and in the format of the CAS Mathematica. LaTeX is a common language for encoding the presentation of mathematical formulae. In contrast to LaTeX, Mathematica's representation focuses on making formulae computable. Hence the content must be encoded, i.e., both the structure and the semantics of mathematical formulae must be taken into consideration.

In LaTeX, the Riemann hypothesis can be expressed using the following string:

\noindent\verb|\zeta(s) = 0 \Rightarrow \Re s = \frac12 \lor \Im s=0|.

\noindent In Mathematica, the Riemann hypothesis can be represented as:

\noindent\texttt{Implies[Equal[Zeta[s], 0], Or[Equal[Re[s], Rational[1, 2]], Equal[Im[s], 0]]]}.

\delimitershortfall=-.2pt%growing brackets
\fontdimen16\textfont2=3pt%larger space down
\fontdimen17\textfont2=3pt
\newcounter{tok}
\DeclareRobustCommand{\token}[4][]{\refstepcounter{tok}%
{{#2}}_{\text{#3}}^%
{\mathbin{\textcolor{gray}{\scriptscriptstyle\arabic{tok}}}}%\ifthenelse{\equal{#1}{}}{}{%\label{#1}
}
\noindent 

The conversion between these two formats is challenging due to a range of conceptual and technical differences. 

First, the grammars underlying the two representation formats greatly differ. LaTeX uses the unrestricted grammar of the TeX typesetting system. The entire set of commands can be re-defined and extended at runtime, which means that TeX effectively allows its users to change every character used for the markup, including the \textbackslash\ character typically used to start commands. The large degree of freedom of the TeX grammar significantly complicates recognizing even the most basic tokens contained in mathematical formulae. In difference to LaTeX, CAS use a significantly more restrictive grammar consisting of a predefined set of keywords and set rules that govern the structure of expressions. For example in Mathematica, function arguments must always be enclosed in square brackets and separated by commas.

Second, the extensive differences in the grammars of the two languages are reflected in the resulting expression trees. Similar to parse trees in natural language, the syntactic rules of mathematical notation, such as operator precedence and function scope, determine a hierarchical structure for mathematical expressions that can be understood, represented, and processed as a tree. The mathematical expression trees of formulae consist of functions or operators and their arguments. We used nested square brackets to denote levels of the tree and Arabic numbers in a gray font to indicate individual tokens in the markup. For the LaTeX representation of the Riemann hypothesis, the expression tree is:  
\begin{align}
\left[%\token{}{e}{1}\,
  \token{\zeta}{l}{1}\,
  \token{(}{l}{1}\,
  \token{s}{l}{1}\,
  \token{)}{l}{1}\,
  \token{=}{l}{1}\,
  \token{0}{l}{1}\,
  \token[pl.impl]{\Rightarrow}{l}{1}\,
  \token{\Re}{l}{1}\,
  \token{s}{l}{1}\,
  \token{=}{l}{1}\,
   \left[\token{}{$\tfrac{\cdot}{\cdot}$}{1}\,
    \token{1}{l}{2}\,
    \token{2}{l}{2}
  \right]\,
  \token{\lor}{l}{1}\,
  \token{\Im}{l}{1}\,
  \token{s}{l}{1}\,
  \token{=}{l}{1}\,
  \token{0}{l}{1}
\right].\nonumber
\end{align}
The tree consists of 18 nodes, i.e., tokens, with a maximum depth of two (for the fraction command \verb \frac12 ). The expression tree of the Mathematica expression consists of 16 tokens with a maximum depth of five:
\begin{align} 
\left[\token[cl.impl]{}{$\Rightarrow$}{1}%
  \left[\token{}{$=$}{2}%
    \left[\token{}{$\zeta$}{3}
      \token{s}{l}{4}
    \right]\,
    \token{0}{n}{3}%
  \right]\,
  \left[\token{}{$\lor$}{2}
    \left[\token{}{$=$}{3}
      \left[\token{}{$\Re$}{4}
        \token{s}{l}{5}
      \right]\,
      \left[\token{}{$\mathbb{Q}$}{4}
        \token{1}{n}{5}\,
        \token{2}{n}{5}
      \right]\,
    \right]\,
    \left[\token{}{$=$}{3}
      \left[\token{}{$\Im$}{4}
        \token{s}{l}{5}
      \right]\,
      \token{0}{n}{4}
    \right]\,
  \right]
\right].\nonumber
\end{align}
The higher complexity of the Mathematica expression reflects that a CAS represents the content structure of the formula, which is deeply nested. In contrast, LaTeX exclusively represents the presentational layout of the Riemann hypothesis, which is almost linear.   

For the given example of the Riemann hypothesis, finding alignments between the tokens in both representations and converting one representation into the other is possible. In fact, Mathematica and other CAS offer a direct import of TeX expressions, which we evaluate in Section~\ref{sc.evalI}. 

However, aside from technical obstacles, such as reliably determining tokens in TeX expressions, conceptual differences also prevent a successful conversion between presentation languages, such as TeX, and content languages. Even if there was only one generally accepted presentation language, e.g., a standardized TeX dialect, and only one generally accepted content language, e.g., a standardized input language for CAS, an accurate conversion between the representation formats could not be guaranteed. 

The reason is that neither the presentation language, nor the content language always provides all required information to convert an expression to the respective language. This can be illustrated by the simple expression: $F(a+b)= Fa +Fb$. The inherent content ambiguity of $F$ prevents a deterministic conversion from the presentation language to a content language. $F$ might, for example, represent a number, a matrix, a linear function or even a symbol. Without additional information, a correct conversion to a content language is not guaranteed. On the other hand, the transformation from content language to presentation language often depends on the preferences of the author and the context. For example, authors sometimes change the presentation of a formula to focus on specific parts of the formula or improve its readability.

Another obstacle to conversions between typical presentation languages and typical content languages, such as the formats of CAS, are the restricted set of functions and the simpler grammars that CAS offer. While TeX allows users to express the presentation of virtually all mathematical symbols, thus denoting any mathematical concept, CAS do not support all available mathematical functions or structures. A significant problem related to the discrepancy of the space of concepts expressible using presentation markup and the implementation of such concepts in CAS are branch cuts. Branch cuts are restrictions of the set of output values that CAS impose for functions that yield ambiguous, i.e., multiple mathematically permissible outputs. One example is the complex logarithm~\cite[eq. 4.2.1]{DLMF}, which has an infinite set of permissible outputs resulting from the periodicity of its inverse function. To account for this circumstance, CAS typically restrict the set of permissible outputs by cutting the complex plane of permissible outputs. However, since the method of restricting the set of permissible outputs varies between systems, identical inputs can lead to drastically different results~\cite{Cohl17}. For example, multiple scientific publications address the problem of accounting for branch cuts when entering expressions in CAS, such as~\cite{england2014branch} for MAPLE.

Our review of obstacles to the conversion of representation formats for mathematical formulae highlights the need to store \textit{both} presentation and content information to allow for reversible transformations. Mathematical representation formats that include presentation and content information can enable the reliable exchange of information between typesetting systems and CAS. 

\sloppy MathML offers standardized markup functionality for both presentation and content information. Moreover, the declarative MathML XML format is relatively easy to parse and allows for cross references between presentation language (PL) and content language (CL) elements. Listing \ref{lst.MathML} represents excerpts of the MathML markup for our example of the Riemann hypothesis (\ref{eq.rh}). In this excerpt, the PL token 7 corresponds to the CL token 19, PL token 5 corresponds to CL token 20, and so forth.

\begin{lstlisting}[label={lst.MathML},mathescape=true,float,caption=MathML representation of the Riemann hypothesis (\ref{eq.rh}) (excerpt).]
<math><semantics><mrow>$\dots$
  <mo id="5" xref="20">=</mo>
  <mn id="5" xref="21">0</mn>
  <mo id="7" xref="19">$\Rightarrow$</ci>$\dots$</mrow>
<annotation-xml encoding="MathML-Content">
  <apply><implies id="19" xref="7"/>
  <apply><eq id="20" xref="5"/>$\dots$
  <apply><csymbol id="21" xref="1" cd="wikidata">$\href{https://www.wikidata.org/w/index.php?title=Q187235&oldid=616744815}{Q187235}\dots$
</annotation-xml></semantics></math>
\end{lstlisting}

\fussy Combined presentation and content formats, such as MathML, significantly improve the access to mathematical knowledge for users of digital libraries.  For example, including content information of formulae can advance search and recommendation systems for mathematical content. The quality of these \textit{mathematical information retrieval systems} crucially depends on the accuracy of the computed document-query and document-document similarities. Considering the content information of mathematical formulae can improve these computations by:
\begin{enumerate}
\item enabling the consideration of mathematical equivalence as a similarity feature. Instead of exclusively analyzing presentation information as indexed, e.g., by considering the overlap in presentational tokens, content information allows modifying the query and the indexed information. For example, it would become possible to recognize that the expressions $a(\frac{b}{c} + \frac{d}{c})$ and $\frac{a(b+d)}{c}$ have a distance of zero.
\item allowing the association of mathematical tokens with mathematical concepts. For example, linking identifiers, such as $E$, $m$, and $c$, to energy, mass, and speed of light, could enable searching for all formulae that combine all or a subset of the concepts. 
\item enabling the analysis of structural similarity. The availability of content information would enable the application of measures, such as derivatives of the tree edit distance, to discover structural similarity, e.g., using $\lambda$-calculus. This functionality could increase the capabilities of \textit{math-based plagiarism detection systems} when it comes to identifying obfuscated instances of reused mathematical formulae \cite{Meuschke2017a}.  
\end{enumerate}

Content information could furthermore enable interactive support functions for consumers and producers of mathematical content. For example, readers of mathematical documents could be offered interactive computations and visualizations of formulae to accelerate the understanding of STEM documents. Authors of mathematical documents could benefit from automated editing suggestions, such as auto completion, reference suggestion, and sanity checks, e.g., type and definiteness checking, similar to the functionality of word processors for natural language texts.

\subsection*{Related Work}\label{sc.relW}
A variety of tools exist to convert format representations of mathematical formulae. However, to our knowledge, Kohlhase et al. presented the only study that evaluated the conversion quality of tools ~\cite{stamerjohanns2009mathml}. Unfortunately, many of the tools evaluated by Kohlhase et al.~are no longer available or out of date.
Watt presents a strategy to preserve formula semantics in TeX to MathML conversions. His approach relies on encoding the semantics in custom TeX
macros rather than to expand the macros~\cite{watt2002exploiting}.
Padovani discusses the roles of MathML and TeX elements for managing large repositories of mathematical knowledge ~\cite{DBLP:conf/mkm/Padovani03}.
Nghiem et al. used statistical machine translation to convert presentation to content language \cite{nghiem2012automatic}.
However, they do not consider the textual context of formulae.
We will present detailed descriptions and evaluation results for specific conversion approaches in Section~\ref{sc.evalI}. 

\citeauthor{Youssef2017} addressed the semantic enrichment of mathematical formulae in presentation language. They developed an automated tagger that parses LaTeX formulae and annotates recognized tokens very similarly to Part-Of-Speech (POS) taggers for natural language~\cite{Youssef2017}. Their tagger currently uses a predefined, context-independent dictionary to identify and annotate formula components. \citeauthor{Schubotz2017} proposed an approach to semantically enrich formulae by analyzing their textual context for the definitions of identifiers~\cite{Schubotz2017,disSigir16}.

With their `math in the middle approach', Dehaye et al. envision an entirely different approach to exchanging machine readable mathematical expressions. In their vision, independent and enclosed virtual research environments use a standardized format for mathematics to avoid computions and transfers between different systems.~\cite{Dehaye2016}.

For an extensive review of format conversion and retrieval approaches for mathematical formulae, refer to ~\cite[Chapter 2]{dis}. 

\section{Benchmarking MathML} \label{sc.gold}This section presents MathMLben - a benchmark dataset for measuring the quality of MathML markup of mathematical formulae appearing in a textual context. MathMLben is an improvement of the gold standard provided by \citeauthor{disSigir15}~\cite{disSigir15}. The dataset considers recent discussions of the \href{http://imkt.org/}{Intrenational Mathematical Knowledge of Trust} working group, in particular the idea of a `Semantic Capture Language'~\cite{IonW17}, which makes the gold standard more robust and easily accessible. MathMLben: 
\begin{itemize}
\item allows comparisons to prior works;
\item covers a wide range of research areas in STEM literature;
\item provides references to manually annotated and corrected MathML items that are compliant with the MathML standard;
\item is easy to modify and extend, i.e., by external collaborators;
\item includes default distance measures; and
\item facilitates the development of converters and tools.
\end{itemize}

In Section~\ref{sc.data}, we present the test collection included in MathMLben.
In Section~\ref{sc.encoding}, we present the encoding guidelines for the human assessors and describe the tools we developed to support assessors in creating the gold standard dataset.
In Section~\ref{sc.sim}, we describe the similarity measures used to assess the markup quality.

\subsection{Collection}\label{sc.data}
Our test collection contains 305 formulae (more precisely, mathematical expressions ranging from individual symbols to complex multi-line formulae) and the documents in which they appear. 

{\bf Expressions 1 to 100} correspond to the search targets used for the `National Institute of Informatics Testbeds and Community for Information access Research Project' (NTCIR) 11 Math Wikipedia Task~\cite{disSigir15}. This list of formulae has been used for formula search and content enrichment tasks by at least 7 different research institutions. The formulae were randomly sampled from Wikipedia and include expressions with incorrect presentation markup.

{\bf Expressions 101 to 200} are random samples taken from the NIST Digital Library of Mathematical Functions (DLMF)~\cite{DLMF}.
The DLMF website contains 9,897 labeled formulae created from semantic LaTeX source files~\cite{disCicm14Drmf,disCicm15}. %N: @AAA / @MMM I deleted "taken in a recent time slice", because I didn't understand what was meant ... put back in if important, but explain it better in that case 
In contrast to the examples from Wikipedia, all these formulae are from the mathematics research field and exhibit high quality presentation markup. The formulae were curated by renowned mathematicians and the editorial board keeps improving the quality of the formulae's markup\footnote{\url{http://dlmf.nist.gov/about/staff}}. 
Sometimes, a labeled formula contains multiple equations. In such cases, we randomly chose one of the equations.

{\bf Expressions 201 to 305} were chosen from the queries of the NTCIR arXiv and NTCIR-12 Wikipedia datasets.
70\% of these queries originate from the arXiv~\cite{disNtcir11Ov} and 30\% from a Wikipedia dump.

All data is openly available for research purposes and can be obtained from: \url{https://mathmlben.wmflabs.org}\footnote{Visit \url{https://mathmlben.wmflabs.org/about} for a user guide.}.

\subsection{Gold Standard}\label{sc.encoding}

We provide explicit markup with universal, context-independent symbols in content MathML.
Since the symbols from the default content dictionary of MathML\footnote{\url{http://www.openmath.org/cd}} alone were insufficient to cover the range of semantics in our collection, we added the Wikidata content dictionary~\cite{schubotz16implCd}. As a result, we could refer to all Wikidata items as symbols in a content tree. This approach has several advantages. Descriptions and labels are available in many languages. Some symbols even have external identifiers, e.g., from the Wolfram Functions Site, or from stack-exchange topics. All symbols are linked to Wikipedia articles, which offer extensive human-readable descriptions. Finally, symbols have relations to other Wikidata items, which opens a range of new research opportunities, e.g., for improving the taxonomic distance measure~\cite{disNtcir11Sim}. 

Our Wikidata-enhanced, yet standard-compliant MathML markup, facilitates the manual creation of content markup. To further support human assessors in creating content annotations, we extended the VMEXT visualization tool~\cite{vmext17} to develop a visual support tool for creating and editing the \textit{MathMLBen} gold standard.

\begin{table}[]
\centering
\caption{Special content symbols added to LaTeXML for the creation of the gold standard.}
\label{tb.symbols}
\begin{tabular}{llll}
\textbf{No} & \textbf{rendering}                 & \textbf{meaning}              & \textbf{example ID} \\
1           & $\left[\qvar{x},\qvar{y}\right]$   & \w{2989763}{commutator}       & \qId{91}            \\
2           & ${\qvar{x}}^{\qvar{y}}_{\qvar{z}}$ & \w{188524}{tensor}            & \qId{43}, \qId{208}, \qId{226}\\
3           & ${\qvar{x}}^{\dagger}$             & \w{2051983}{adjoint}          & \qId{224}, \qId{277}\\
4           & ${\qvar{x}}^{'}$                   & \w{Q12202238}{transformation} & \qId{20}            \\
5           & ${\qvar{x}}^{\circ}$               & \w{Q28390}{degree}            & \qId{20}            \\
6			& $\qvar{x}^{(dim)}$				 & \w{Q5165685}{contraction}     & \qId{225}
\end{tabular}
\end{table}

For each formula, we saved the source document written in different dialects of LaTeX and converted it into content MathML with parallel markup using LaTeXML~\cite{LaTeXML,DBLP:conf/lwa/GinevSK11}. LaTeXML is a Perl program that converts LaTeX documents to XML and HTML. We chose LaTeXML, because it is the only tool that supports our semantic macro set. We manually annotated our dataset, generated the MathML representation, manually corrected errors in the MathML, and linked the identifiers to Wikidata concept entries whenever possible. Alternatively, one could initially generate MathML using a CAS and then manually improve the markup.

Since there is no generally accepted definition of expression trees, we made several design decision to create semantic representations of the formulae in our dataset using MathML trees. In some cases, we created new macros to be able to create a MathML tree for our purposes using LaTeXML\footnote{\url{http://dlmf.nist.gov/LaTeXML/manual/customization/customization.latexml.html\#SS1.SSS0.Px1}}. Table~\ref{tb.symbols} lists the newly created macros. Hereafter, we explain our decisions and give examples of formulae in our dataset that were affected by the decisions. 

\begin{itemize}
\item not assign Wikidata items to basic mathematical identifiers and functions like \verb|factorial|, \verb|\log|, \verb|\exp|, \verb|\times|, \verb|\pi|. Instead, we left these annotations to the DLMF LaTeX macros, because they represent the mathematical concept by linking to the definition in the DLMF and LaTeXML creates valid and accurate content MathML for these macros [GoldID \qId{3}, \qId{11}, \qId{19}, ...];
\item split up indices and labels of elements as child nodes of the element. For example, we represent \verb|i| as a child node of \verb|p| in \verb|p_i| [GoldID \qId{29}, \qId{36}, \qId{43}, ...];
\item create a special macro to represent tensors, such as for $T_{\alpha\beta}$ [GoldID \qId{43}], to represent upper and lower indices as child nodes (see table~\ref{tb.symbols});
\item create a macro for dimensions of tensor contractions [GoldID \qId{225}], e.g., to distinguish the three dimensional contraction of the metric tensor in $g^{(3)}$ from a power function (see table~\ref{tb.symbols});
\item chose one subexpression randomly if the original expression contained lists of expressions [GoldID \qId{278}];
\item remove equation labels, as they are not part of the formula itself. For example, in
\begin{equation}\label{eq:starEx}
	E = mc^2, \tag{$\star$}
\end{equation}
the (\ref{eq:starEx}) is the ignored label;
\item remove operations applied to entire equations, e.g., applying the modulus. In such cases, we interpreted the modulus as a constraint of the equation [GoldID \qId{177}];
\item use additional macros (see table~\ref{tb.symbols}) to interpret complex conjugations, transformation signs, and degree-symbols as functional operations (identifier is a child node of the operation symbol), e.g., \verb|*| or \verb|\dagger| for complex conjugations [GoldID \qId{224}, \qId{277}], \verb|S'| for transformations [GoldID \qId{20}], \verb|30^\circ| for thirty degrees [Gold ID 30];
\item for formulae with multiple cases, render each case as a separate branch [GoldID \qId{49}];
\item render variables that are part of separate branches in bracket notation. We implemented the Dirac Bracket commutator [] (omitting the index \verb|_\text{DB}|) and an anticommutator {} by defining new macros (see table \ref{tb.symbols}). Thus, there is a distinction between a (ring) commutator \verb|[a,b] = ab - ba| and an anticommutator \verb|{a,b} = ab + ba|, without further annotation of Dirac or Poisson brackets [GoldID \qId{91}];
\item use the command \verb|\operatorname{}| for multi-character identifiers or operators [GoldID \qId{22}]. This markup is necessary, because most LaTeX parsers, including LaTeXML, interpret multi-character expressions as multiplications of the characters. In general, this interpretation is correct, since it is inconvenient to use multi-character identifiers~\cite{cajori1928history}.
\end{itemize}

Some of these design decisions are debatable. For example, introducing a macro \verb|\identifiername{}| to distinguish between multi-character identifiers and operators might be advantageous to our approach. However, introducing many highly specialized macros is likely not a viable approach and exaggerated. A borderline example in regard to this problem is $\Delta x$ [GoldID \qId{280}]. Formulae of this form could be annotated as \verb|\operatorname{}|, \verb|\identifiername{}| or more generally as \verb|\expressionname{}|. We interpret $\Delta$ as a difference applied to a variable, and render the expression as a function call.

Similar cases of overfeeding the dataset with highly specialized macros are bracket notations. For example, the bracket (Dirac) notation, e.g., [GoldID \qId{209}], is mainly used in quantum physics. The angle brackets for the Dirac notation, $\langle$ and $\rangle$, and a vertical bar | is already interpreted correctly as "latexml - quantum-operator-product". However, a more precise distinction between a twofold scalar product, e.g., $\langle a|b\rangle$, and a threefold expectation value, e.g., $\langle a|A|a\rangle$, might become necessary in some scenarios to distinguish between matrix elements and a scalar product.

\begin{figure*}[t]
\includegraphics[scale=0.45]{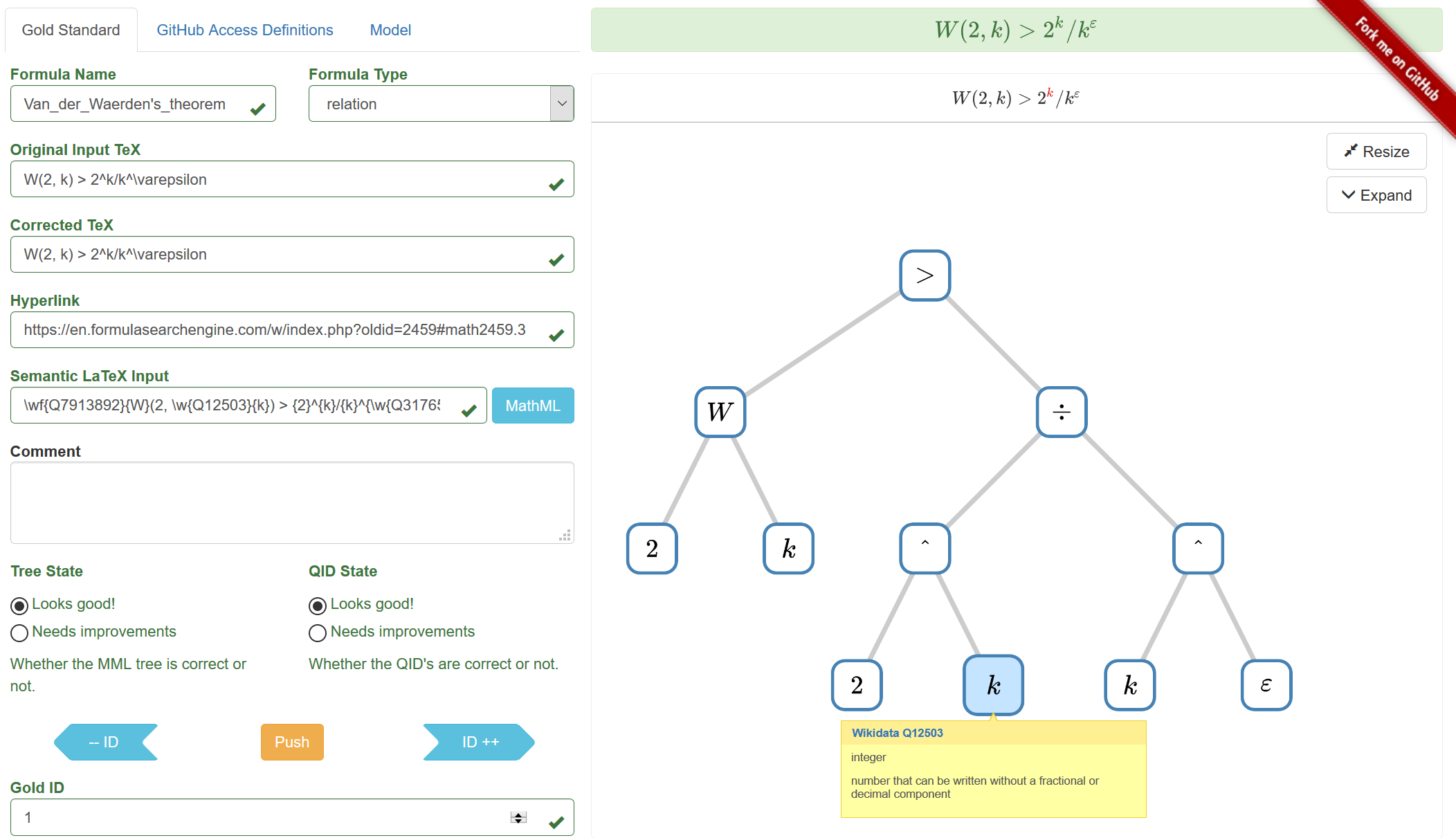}
\caption{Graphical User Interface to support the creation of our gold standard. The interface provides several TeX input fields (left) and a mathematical expression tree rendered by the VMEXT visualization tool (right).}
\label{fig:GouldiGUI}
\end{figure*}

We developed a Web application to create and cultivate the gold standard entries, which is available at: \url{https://mathmlben.wmflabs.org/}. The Graphical User Interface (GUI) provides the following information for each Gold ID entry.\\[-24pt]

\begin{itemize}
\item \textbf{Formula Name:} the name of the formula (optional)
\item \textbf{Formula Type:} either \textit{definition}, \textit{equation}, \textit{relation} or \textit{General Formula} (if none of the previous names fit)
\item \textbf{Original Input TeX:} the LaTeX expression extracted from the source
\item \textbf{Corrected TeX:} the manually corrected LaTeX expression
\item \textbf{Hyperlink:} the hyperlink to the position of the formula in the source
\item \textbf{Semantic LaTeX Input:} the manually created semantic version of the corrected LaTeX field. This entry is used to generate our MathML with Wikidata annotations.
\item \textbf{Preview of Corrected LaTeX:} a preview of the corrected LaTeX input field rendered as an SVG image in real time using Mathoid~\cite{Mathoid}, a service to generate SVGs and MathML from LaTeX input. It is shown in the top right corner of the GUI.
\item \textbf{VMEXT Preview:} the VMEXT field renders the expression tree based on the content MathML. The symbol in each node is associated with the symbol in the cross referenced presentation MathML.
\end{itemize}

Figure~\ref{fig:GouldiGUI} shows the GUI that allows to manually modify the different formats of a formula. While the other fields are intended to provide additional information, the pipeline to create and cultivate a gold standard entry starts with the semantic LaTeX input field. LaTeXML will generate content MathML based on this input and VMEXT will render the generated content MathML afterwards. We control the output by using the DLMF LaTeX macros~\cite{DBLP:journals/amai/MillerY03} and our developed extensions. The following list contains some example of the DLMF LaTeX macros.

\begin{itemize}
\setlength\itemsep{0.15cm}
\item \verb|\EulerGamma@{z}|: $\Gamma(z)$: gamma function,
\item \verb|\BesselJ{\nu}@{z}|: $J_\nu(z)$: Bessel function of the first kind,
\item \verb|\LegendreQ[\mu]{\nu}@{z}|: $Q^\mu_\nu(z)$:\\ associated Legendre function of the second kind,
\item \verb|\JacobiP{\alpha}{\beta}{n}@{x}|: $P^{(\alpha,\beta)}_n(x)$:\\ Jacobi polynomial.
\end{itemize}

The DLMF web pages, which we use as one of the sources for our dataset, were generated from semantically enriched LaTeX sources using LaTeXML. Since LaTeXML is capable to interpret semantic macros, generates content MathML that can be controlled with macros, and is easily extensible by new macros, we also used LaTeXML to generate our gold standard. While the DLMF is a compendium for special functions, we need to annotate every identifier in the formula with semantic information. Therefore, we extended the set of semantic macros.

In addition to the special symbols listed in Table~\ref{tb.symbols}, we created macros to semantically enrich identifiers, operators, and other mathematical concepts by linking them to their Wikidata items. As shown in Figure \ref{fig:GouldiGUI}, the annotations are visualized using yellow info boxes appearing on mouse over. The boxes show the Wikidata QID, the name, and the description (if available) of the linked concept.

Aside from naming, classifying, and semantically annotating each formula, we performed three other tasks:

\begin{itemize}
\setlength\itemsep{0.15cm}
\item correcting the LaTeX string extracted from the sources;
\item checking and correcting the MathML generated by LaTeXML
\item visualizing the MathMl using VMEXT
\end{itemize}

Most of the extracted formulae contained concepts to improve human readability of the source code, such as commented line breaks, \verb|%\n|, 
in long mathematical expressions, or special macros to improve the displayed version of the formula, e.g., spacing macros, delimiters, and scale settings, such as \verb|\!|, \verb|\,| or \verb|\>|. Since they are part of the expression, all of the tested tools (also LaTeXML) try to include these formating improvements into the MathML markup. For our gold standard, we focus on the pure semantic information and forgo formating improvements related to displaying the formula. The corrected TeX field shows the cleaned mathematical LaTeX expression.

Using the corrected TeX field and the semantic macros, we were able to adjust the MathML output using LaTeXML and verify it by checking the visualization from VMEXT.

\subsection{Evaluation Metrics}\label{sc.sim}

To quantify the conversion quality of individual tools, we computed the similarity of each tool's output and the manually created gold standard. To define the similarity measures for this comparison, we built upon our previous work~\cite{disNtcir11Sim}, in which we defined and evaluated four similarity measures: taxonomic distance, data type hierarchy level, match depth, and query coverage. The measures taxonomic distance and data type hierarchy level require the availability of a hierarchical ordering of mathematical functions and objects. For our use case, we derived this hierarchical ordering from the MathML content dictionary. The measures assign a higher similarity score if matching formula elements belong to the same taxonomic class. The match depth measure operates under the assumption that matching elements, which are more deeply nested in a formula's content tree, i.e., farther away from the root node, are less significant for the overall similarity of the formula, hence are assigned a lower weight. The query coverage measure performs a simple `bag of tokens' comparison between two formulae and assigns a higher score the more tokens the two formulae share.

In addition to these similarity measures, we also included the tree edit distance. For this purpose, we adapted the robust tree edit distance (RTED) implementation for Java~\cite{RTED}.
We modified RTED to accept any valid XML input and added math-specific `shortcuts', i.e., rewrite rules that generate lower distance scores than arbitrary rewrites.
For example, rewriting $\frac{a}{b}$ to $ab^{-1}$ causes a significant difference in the expression tree:
Three nodes ($\wedge,-,1$) are inserted and one node is renamed $\div\to\cdot.$
The `costs' for performing these edits using the stock implementation of RTED are $c=3i+r.$
However, the actual difference is an equivalence, which we think should be assigned a cost of $e<3i+r.$
We set $e<r<i.$

\section{Evaluation Context-agnostic Conversion Tools} \label{sc.evalI}
This section presents the results of evaluating existing, context-agnostic conversion tools for mathematical formulae using our benchmark dataset MathMLben (cf.~Section~\ref{sc.gold}). We compare the distances between the presentation MathML and the content MathML tree of a formula yielded by each tool to the respective trees of formulae in the gold standard. We use the tree edit distance with customized weights and math-specific shortcuts. The goal of shortcuts is eliminating notational-inherent degrees of freedom, e.g., additional PL elements or layout blocks, such as \verb|mrow| or \verb|mfenced|. %See also \href{https://arxiv.org/pdf/1103.1252.pdf}{ted for html}.

\subsection{Tool Selection}
We compiled a list of available conversion tools from the W3C\footnote{\url{https://www.w3.org/wiki/Math_Tools}} wiki, from \textit{GitHub}, and from questions about automated conversion of mathematical LaTeX to MathML on \textit{Stack Overflow}. 
We selected the following converters:

\begin{itemize}
\item LaTeXML: can convert generic and semantically annotated LaTeX expressions to XML/HTML/MathML. The tool is written in Perl~\cite{LaTeXML} and is actively maintained. LaTeXML was specifically developed to generate the DLMF web page and can therefore parse entire TeX documents. LaTeXML also supports conversions to content MathML.
\item LaTeX2MathML: is a small python project and is able to generate presentation MathML from generic LaTeX expressions\footnote{\url{https://github.com/Code-ReaQtor/latex2mathml}}.
\item Mathoid: is a service developed using Node.js, PhantomJS and MathJax (a javascript display engine for mathematics) to generate SVGs and MathML from LaTeX input. Mathoid is currently used to render mathematical formulae on Wikipedia~\cite{Mathoid}.
\item SnuggleTeX: is an open-source Java library developed at the University of Edinburgh\footnote{\url{https://www2.ph.ed.ac.uk/snuggletex/documentation/overview-and-features.html}}. The tool allows to convert simple LaTeX expression to XHTML and presentation MathML.
\item MathToWeb: is an open-source Java-based web application that generates presentation MathML from LaTeX expressions\footnote{\url{https://www.mathtowebonline.com}}.
\item TeXZilla: is a javascript web application for LaTeX to MathML conversion capable of handling Unicode characters\footnote{\url{https://fred-wang.github.io/TeXZilla}}.
\item Mathematical: is an application written in C and wrapped in Ruby to provide a fast translation from LaTeX expressions to the image formats SVG and PNG. The tool also provides translations to presentation MathML\footnote{\url{https://github.com/gjtorikian/mathematical}}.
\item CAS: we included a prominent CAS capable of parsing LaTeX expressions.
\item Part-of-Math (POM) Tagger: is a grammar-based LaTeX parser that tags recognized tokens with information from a dictionary~\cite{Youssef2017}. The POM tagger is currently under development. In this paper, we use the first version. In~\cite{Cohl17}, this version was used to provide translations LaTeX to the CAS MAPLE. In its current state, the program offers no export to MathML. We developed an XML exporter to be able to compare the tree provided by the POM tagger with the MathML trees in the gold standard.
\end{itemize}

\subsection{Testing framework}
We developed a Java-based framework that calls the programs to parse the corrected TeX input data from the gold standard to presentation MathML, and, if applicable, to content MathML. In case of the POM tagger, we parsed the input string to a general XML document. We used the corrected TeX input format instead of the originally extracted string expressions, see~\ref{sc.encoding}.

Executing the testing framework requires the manual installation of the tested tools. The POM tagger is not yet publicly available. 

\subsection{Results}

\begin{figure}[t]
\includegraphics[trim={6cm 2.2cm 6cm 2.2cm},clip,width=\linewidth]{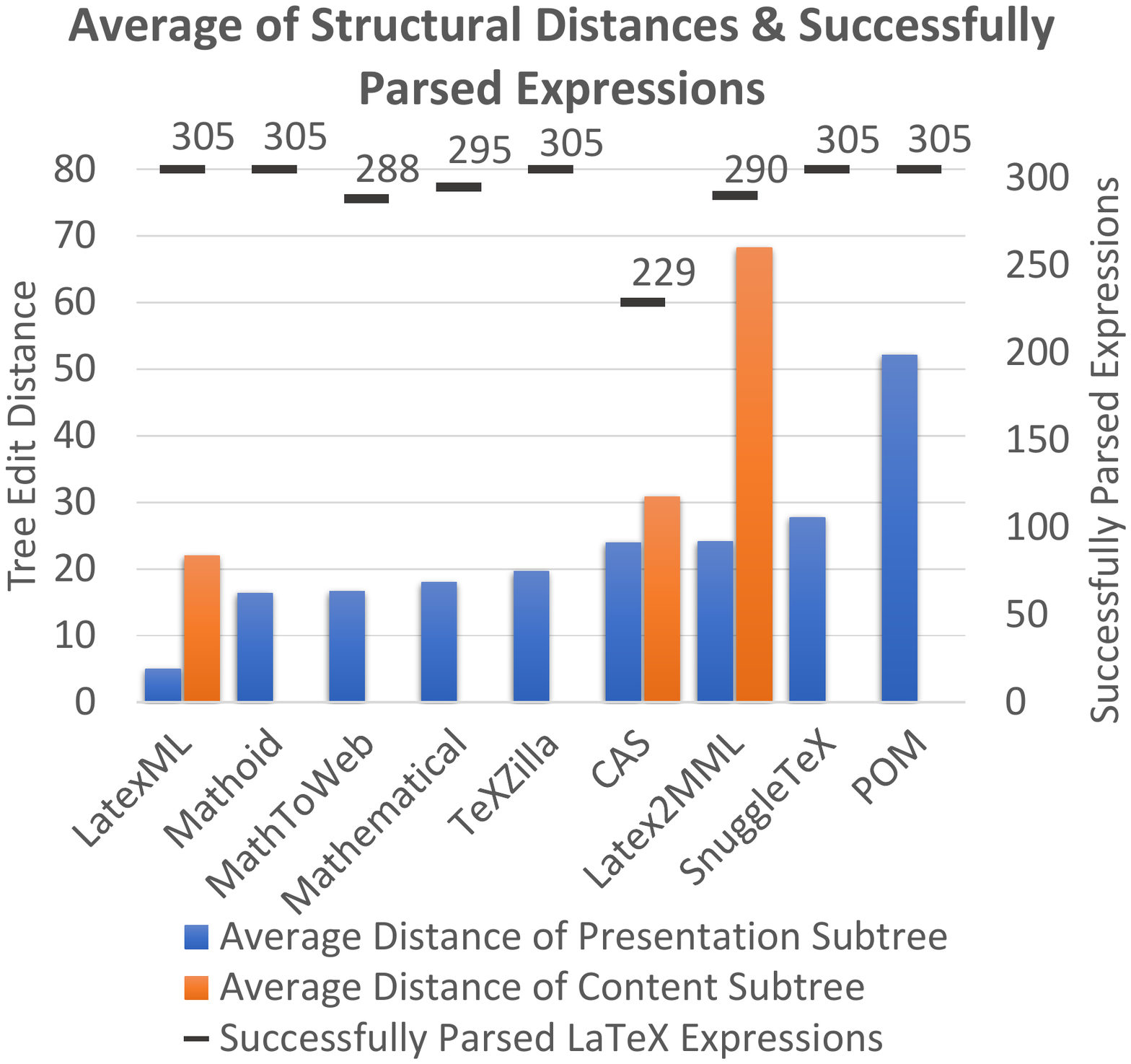}
\caption{Overview of the structural tree edit distances (using $r=0,\ i=d=1$) between the MathML trees generated by the conversion tools and the gold standard MathML trees.}
\label{fig:results}
\vspace*{-0.7em}
\end{figure}

\begin{figure}[t]
\includegraphics[trim={4cm 2cm 4cm 2cm},clip,width=\linewidth]{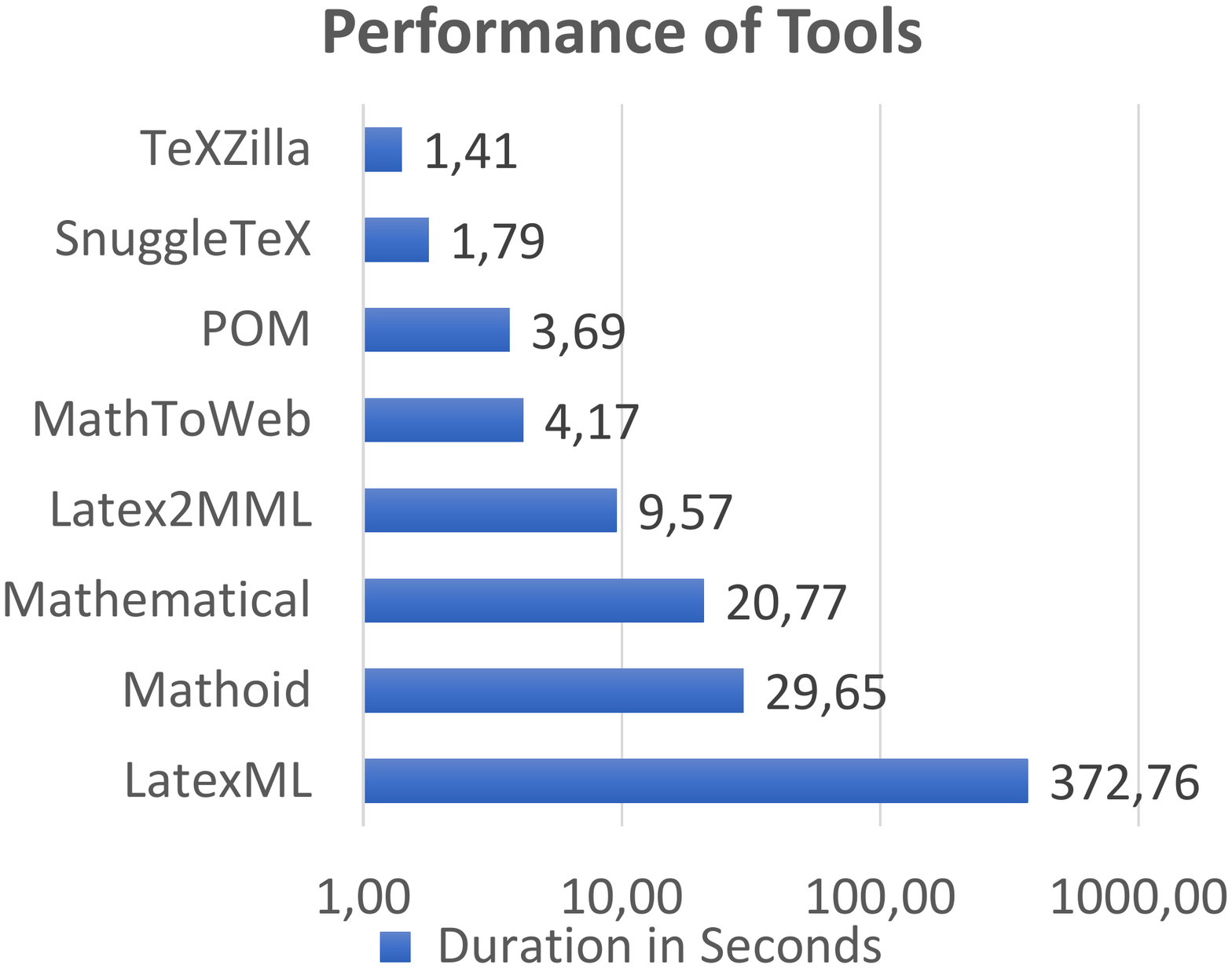}
\caption{Time in seconds required by each tool to parse the 305 gold standard LaTeX expressions in logarithmic scale.}
\label{fig:performance}
\vspace*{-0.7em}
\end{figure}

Figure~\ref{fig:results} shows the averaged structural tree edit distances between the presentation trees (blue) and content trees (orange) of the generated MathML files and the gold standard. To calculate the structural tree edit distances, we used the RTED~\cite{RTED} algorithm with costs of $i = 1$ for inserting, $d = 1$ for deleting and $r = 0$ for renaming nodes. Furthermore, the Figure shows the total number of successful transformations for the 305 expressions (black ticks). Note that we also consider differences of the presentation tree to the gold standard as deficits, because the mapping from LaTeX expressions to rendered expressions is unique (as long as the same preambles are used). A larger number indicates that more elements of an expression were misinterpreted by the parser. However, certain differences between presentation trees might be tolerable, e.g., reordering commutative expressions, while differences between content trees are more critical. Also note that improving content trees may not necessarily improve presentation trees and vice versa. In case of $f(x+y)$, the content tree will change depending whether $f$ represents a variable or a function, while the presentation tree will be identical in both cases. In contrast, $\frac{a}{b}$, $\sfrac{a}{b}$, and $a/b$ have different presentation trees, while the content trees are identical.

Figure~\ref{fig:performance} illustrates the runtime performance of the tools. We excluded the CAS from the runtime performance tests, because the system is not primarily intended for parsing LaTeX expressions, but for performing complex computations.
Therefore, runtime comparisons between a CAS and conversion tools would not be representative. We measured the times required to transform all 305 expressions in the gold standard and write the transformed MathML to the storage cache. Note that the native code of Latex2MML, Mathematical and LaTeXML were called from the Java Virtual Machine (JVM) and Mathoid was called through local web-requests, which increased the runtime of these tools. The figure is scaled logarithmically. We would like to emphasize that LaTeXML is designed to translate sets of LaTeX documents instead of single mathematical expressions. Most of the other tools are lightweight engines.

In this benchmark, we focused on the structural tree distances rather than on distances in semantics. While our gold standard provides the information necessary to compare the extracted semantic information, we will focus on this problem in future work (see Section~\ref{ssc.futurework}).

\section{Towards a Context-sensitive Approach}\label{sc.appr}
 
\begin{figure*}[h]
\includegraphics[trim={0.5cm 0.5cm 0.5cm 0.5cm},clip,height=0.5\linewidth]{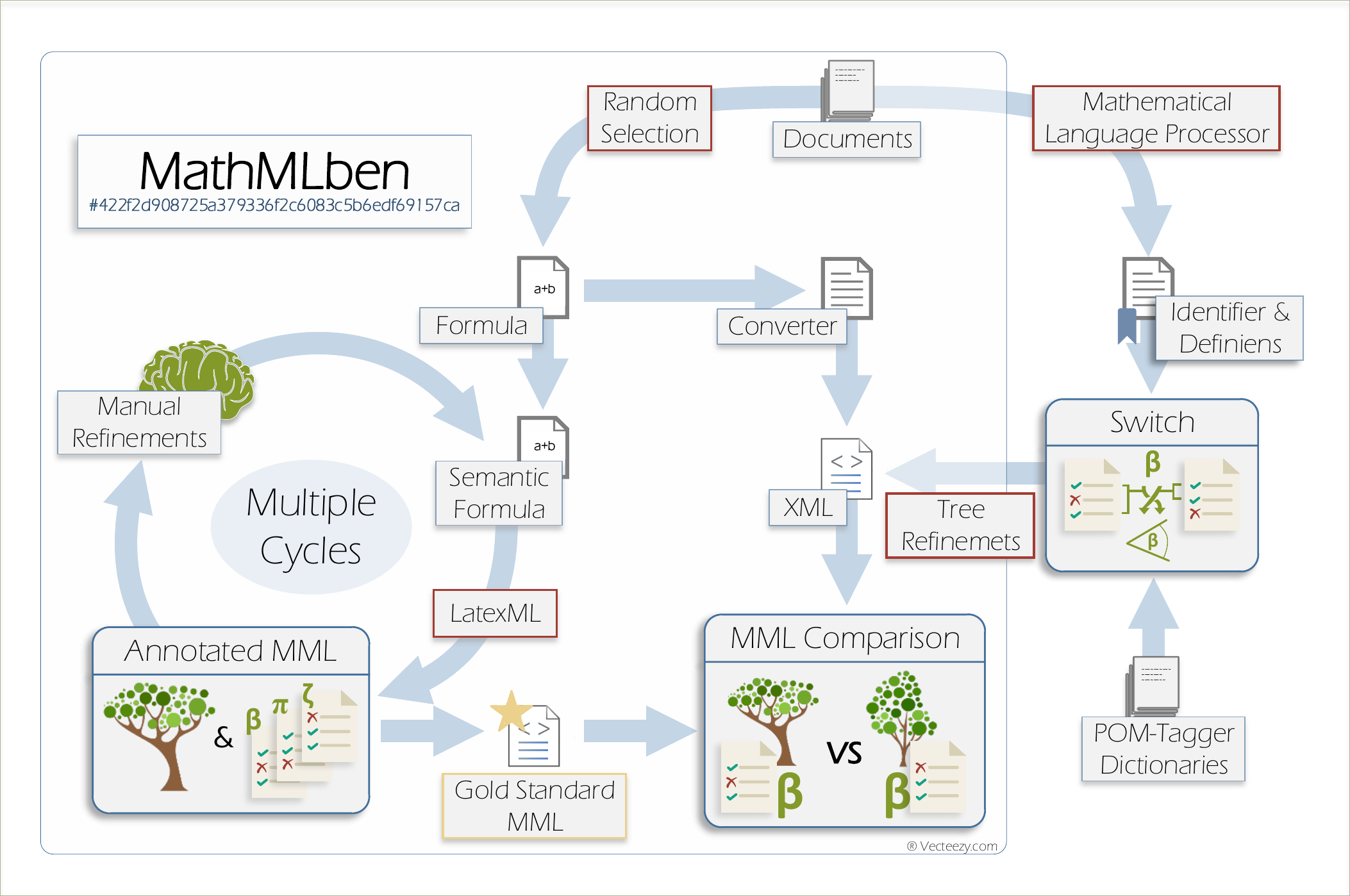}
\caption{Mathematical language processing is the task of mapping textual descriptions to components of mathematical formulae (Part-of-Math tagging).}
\label{fg.overview}
\end{figure*}

In this section, we present our new approach that combines textual features, i.e., semantic information from the surrounding text, with the converters to improve the outcome. 
Figure~\ref{fg.overview} illustrates the process of creating the gold standard, evaluating conversions, and how we plan to improve the converters with tree refinements (outside the MathMLben box).
Our improvement approach includes three phases.
\begin{enumerate}
\item In the first phase, the Mathematical Language Processing (MLP) approach~\cite{disSigir16} extracts semantic information from the textual context by providing identifier-definiens\footnote{In a definition, the \textit{definiendum} is the expression to be defined and \textit{definiens} is the phrase that defines the definiendum. Identifier-definiens pairs are candidates for an Identifier-definition, see~\cite{disSigir16} for a more detailed explanation.} pairs.
\item The MLP annotations self-assess their reliability by annotate each identifier-definiens pair with its probabilities. Often, the methods do not find highly ranked semantic information.
In such cases, we combine the MLP results with a dictionary-based method. In particular, we use the dictionaries from the POM tagger~\cite{Youssef2017} that associate context-free semantics with the presentation tree. Since the dictionary entries are not ranked, we use them to drop unmentioned identifier-definiens pairs and choose the highest rank of the remaining pairs.
\item Based on the chosen semantic information, we redefine the content tree by reordering the nodes and subtrees.
\end{enumerate}

Currently, the implementation is too immature to release it as a semantic annotation package. Instead, we discuss the method using the following selected examples that represent typical classes of disambiguation problems:
\begin{itemize}
\item Invisible operator disambiguation for the times vs.~apply special case.
\item Parameter vs.~label disambiguation for subscripts.
\item Einstein notation discovery.
\item Multi-character operator discovery.
\end{itemize}
Learning special notations like the examples above is subject to future work.
However, we deem it reasonable to start with these examples, since our manual investigation of the tree edit distances showed that such cases represented major reasons for errors in the content MathML tree.
 
Previously, the MLP software was limited to extracting information about identifiers, not general mathematical symbols. Moreover, the software was optimized for the Wikipedia dataset. We thus expanded the software for this study to enable parsing pure XHTML input as provided by the NTCIR tasks and the DLMF website.
Achieving this goal required realizing a component for symbol identification. We chose the strategy of considering every simple expression that is not an identifier as a candidate for a symbol.

For first experiments we tried to improve the output by LaTeXML, since LaTeXML performs best in our tests and it was able to generate content MathML. Moreover, with the newly developed semantic macros, we are able to optimize MathML in a pre-processing step by enhance the input LaTeX expression. In consequence, we do not need to develop complex post-processing algorithms to manipulate content MathML.

As part of this study, we created a custom style sheet that fixes the following problems: (1) use of the power symbols for superscript characters unless Einstein notation was discovered, (2) interpretation of subscript indices as parameters, unless they are in text mode. For text mode, the ensemble of main symbol and subscript will be regarded as an identifier. (3) Symbols that are considered as a `function' are applied to the following identifier, rather than being multiplied with the identifier.

First experiments using the refinement techniques have proven to be very effective. 
We haven chosen a small set of 10 functions for performing the refinements and to show the potential of the techniques. Of those 10 cases, with simple regular expression matching, our MLP approach found 4 cases, where the highest ranked identifier-definiens pair was `function' for at least one identifier in the formula. In these 4 cases, the distances of the content trees decreased to zero with all previously explained refinements enabled.

While this is just a first indication for the suitability of our approach, it shows that the long chain of processing steps shows promise. Therefore, we are actively working on the presented improvements and plan to focus on the task of learning how to generate mappings from the input PL encoding to CL encoding without general rules for branch selection as we applied them so far.

\section{Conclusion and Future Work} \label{sc.concl}
We make available the first benchmark dataset to evaluate the conversion of mathematical formulae between presentation and content formats. During the encoding process for our MathML-based gold standard, we presented the conceptual and technical issues that conversion tools for this task must address. Using the newly created  benchmark dataset, we evaluated popular context-agnostic LaTeX-to-MathML converters. We found that many converters simply do not support the conversion from presentation to content format, and those that did often yielded mathematically incorrect content representations even for basic input data. These results underscore the need for future research on mathematical format conversions.

Of the tools we tested, LaTeXML yielded the best conversion results, was easy to configure, and highly extensible. However, these benefits come at the price of a slow conversion speed. Due to its comparably low error rate, we chose to extend the LaTeXML output with semantic enhancements.

Unfortunately, we failed to develop an automated method to learn special notation.
However, we could show that the application of special selection rules improves the quality of the content tree, i.e., allows choosing the most suitable tree from a selection of candidates.
While the implementation of a few selection rules fixes nearly all issues we encountered in our test documents, the long tail of rules shows the limitations of a rule-based approach.\\[-12px]

\subsubsection*{Future Work}\label{ssc.futurework}
We will focus our future research on methods for automated notation detection, because we consider this approach as better suited and better scalable than implementing complex systems of selection rules. We will extract the considered notational features from the textual context of formulae and use them to extend our previously proposed approach of constructing identifier name spaces~\cite{disSigir16} towards constructing notational name spaces. We will check the integrity of formed notational name spaces with methods comparable to those proposed in our previous publication~\cite{disCicm16Units} where we used physical units as sanity check, if semantic annotation in the domain of physics are correct.

\begin{acks}
We would like to thank Abdou Youssef for sharing his Part-of-Math tagger with us and for offering valuable advice.
We are also indebted to Akiko Aizawa for her advice and for hosting us as visiting researchers in her lab at the National Institute of Informatics (NII) in Tokyo. Furthermore, we thank Wikimedia Labs for providing cloud computing facilities and hosting our gold standard dataset.
This work was supported by the FITWeltweit program of the German Academic Exchange Service (DAAD) as well as the \grantsponsor{DFG}{German Research Foundation (DFG}, grant \grantnum{DFG}{GI-1259-1}).

\end{acks}
\balance
\printbibliography[keyword=primary]
\end{document}